\documentstyle{elsart}
\begin{document}
\begin{frontmatter}

\title{Coexistence of Haldane gap excitations and long-range order in
$R_2$BaNiO$_5$ ($R$=rare earth)}

\author{A. Zheludev}

\address{Physics Department, Brookhaven National Laboratory, Upton, NY
11973-5000, USA}

\begin{abstract}
$R_2$BaNiO$_5$ ($R=$ rare earth)  quasi-1-D antiferromagnets are
structurally equivalent to the well-studied 1-D $S=1$ Haldane-gap
compound Y$_2$BaNiO$_5$. Unlike the Y-nickelate though, these materials
undergo 3-D magnetic ordering at finite temperatures. Recent inelastic
neutron scattering studies of Pr$_2$BaNiO$_5$ and
(Nd$_{x}$Y$_{1-x}$)$_2$BaNiO$_5$ revealed purely 1-dimensional gap
excitations that propagate exclusively on the Ni-chains and are
strikingly similar to Haldane gap  modes in Y$_2$BaNiO$_5$. In the
ordered phase these excitations survive and actually coexist with
conventional spin waves. The results suggest that the Haldane singlet
ground state of the Ni-chains is not fully destroyed by N\'{e}el
ordering.\end{abstract}

\end{frontmatter}


In 1983 the pioneering theoretical work by Haldane\cite{Haldane83} has
radically changed our understanding of low-dimensional (low-D) quantum
magnetic system. It was suggested that the  static and dynamical
properties of a simple 1-D Heisenberg antiferromagnet (AF) are totally
different for integer and half-integer spin models: for {\it integer}
spins the ground state is a non-magnetic singlet, the excitation
spectrum has a {\em gap} and the spin correlations decay exponentially
over a length of only several lattice units. The excitations are a $S=1$
triplet with an energy minimum at the 1-D AF zone-center $Q=\pi
a^{\ast}$, where $2 \pi / a^{\ast}$ is the in-chain lattice constant.
Haldane's conjecture is now supported by a vast amount of theoretical
and numerical results. Most important, in quite a few real 1-D
integer-spin systems the singlet ground state and Haldane gap
excitations have been observed and studied experimentally. Comprehensive
lists of references may be found in several recent publications, as for
example in Refs.\
\cite{Ma92,Kakurai92,Regnault94,Xu96,Zheludev96-NINAZ}.

In the {\it quasi}-1-D case, when magnetic interactions between adjacent
integer-spin chains are finite, we can still expect a quantum-disordered
singlet ground state with a spin gap in the limit of very weak coupling.
For strong inter-chain interactions the system becomes 3-dimensional,
the ground state is N\'{e}el-like, and the excitations are conventional
spin waves with no energy gap. The most physically interesting case is
that of intermediate coupling strengths. A great deal of work  aimed at
understanding this situation was done on CsNiCl$_3$ and related
compounds (see for example Refs.
\cite{Buyers86,Tun91,Affleck-Wellman92,Enderle94} and references
therein). In  the present paper we discuss recent experimental results
obtained for a different family of {\it quasi}-1-D antiferromagnets,
namely those with the general formula $R_{2}$BaNiO$_{5}$, where $R$
stands for one of the rare earth elements or Y. The most important
structural feature of these systems are parallel chains of Ni$^{2+}$
ions running along the $a$ axis of the orthorhombic crystal structure
(See Fig.~1 in Refs.~\cite{Sachan94,Zheludev96PBANO}). The $S=1$
Ni$^{2+}$ ions are linked via the apical oxygen sites of their
coordination octahedra, which results in relatively strong in-chain AF
interactions with $J\approx 25$~meV. The chains are structurally
separated from one another, and the $R^{3+}$ sites are positioned
between the chains. Both Ni-O-$R$ and $R$-O-$R$ superexchange routes may
be active, and in the case when $R^{3+}$ is a magnetic rare earth ion
the system orders antiferromagnetically with $T_{N}$ ranging from 24~K
to almost 80~K
\cite{Alonso90,Garcia93,Garcia95,Butera95,Harker96,Garcia96}. However,
if $R=Y$ the chains are magnetically isolated. No long-range order has
been found so far in Y$_{2}$BaNiO$_{5}$, while Haldane gap excitations
have been observed \cite{Darriet93} and studied in great
detail\cite{Xu96,Sakaguchi96}. $R_{2}$BaNiO$_{5}$ species thus appear to
be almost ideal model materials for studying the cross-over from 1-D
quantum to 3-D classical behavior: i)  in contrast to CsNiCl$_3$, where
the chains form a frustrated 2-D triangular lattice, in
$R_{2}$BaNiO$_{5}$ there is no obvious frustration  of inter-chain
interactions that could complicate the picture; ii) the Haldane gap
$\Delta
\approx 10$~meV in Y$_{2}$BaNiO$_{5}$ is in the energy range readily
accessible in neutron scattering experiments; iii) compounds with
different $R$ substitutes have different magnetic structures and
ordering temperatures, which allows to separate common behaviour from
that characteristic of a particular species; and iv) by preparing
samples with different  proportions of magnetic rare earths and
non-magnetic Y, the experimentalist obtains a {\it direct handle} on the
strength of inter-chain coupling.

In  many ways the $R_{2}$BaNiO$_{5}$ ($R\ne$Y) compounds behave as good
3-D classical antiferromagnets.  Bulk properties and magnetic structures
were investigated for $R$=Pr, Nd, Sm, Eu, Gd, Tb, Dy, Ho, Er and Tm,
using standard magnetic techniques \cite{Garcia95,Butera95,Harker96},
neutron diffraction
\cite{Alonso90,Garcia93,Sachan94,Garcia96,Yokoo97NDY}, and even resonant
magnetic X-ray scattering \cite{Zheludev96NBANOX}. In all compounds the
N\'{e}el temperatures are smaller than the in-chain exchange constant by
less than an order of magnitude. The temperature dependence of magnetic
moments is rather well described by mean-filed models. No deviations of
the order-parameter critical exponent $\beta$ from the Ginsburg-Landau
value $\beta = 0.5$ have been found so far. The only hint of some low-D
and quantum behavior is a suppressed saturated moment of the Ni ions,
typically $1.1~\mu_{B}$, as compared to the classical value of
$2~\mu_{B}$. Two types of magnetic structures with propagation vectors
$(\frac{1}{2},\frac{1}{2},\frac{1}{2})$ exist \cite{Alonso90,Garcia93}.
In Er$_{2}$BaNiO$_{5}$ the spins are almost parallel to the chain
direction ($a$ axis).\cite{Alonso90} In most other materials, such as
Nd$_{2}$BaNiO$_{5}$ and Pr$_{2}$BaNiO$_{5}$, both Ni$^{2+}$ and $R^{3+}$
moments are confined to the $(b,c)$ crystallographic plane
\cite{Sachan94,Zheludev96PBANO,Garcia96}. The structure is non-colinear
in the latter case \cite{Garcia96}, but is roughly as shown in Fig.3 in
Ref.~\cite{Sachan94}. What is important for the following discussion is
that in all  $R_{2}$BaNiO$_{5}$ species both Ni$^{2+}$ and $R^{3+}$
moments order simultaneously, i.e., with a single N\'{e}el temperature.

If $R_{2}$BaNiO$_{5}$  ($R\ne$Y) are, indeed, almost-classical systems,
are the purely quantum-mechanical integer-spin dynamics totally
destroyed? Or do the  spin-singlet state of the Ni chains and the
Haldane-gap excitations somehow survive in the magnetically ordered
species? The main goal of this paper is to argue that it is the latter,
the more physically interesting scenario that is realized. The first
evidence for this were obtained in single-crystal inelastic neutron
scattering experiments on Pr$_{2}$BaNiO$_{5}$ \cite{Zheludev96PBANO}.
The material orders magnetically in a Nd$_{2}$BaNiO$_{5}$-type structure
at $T_{N}=24$~K. Above this temperature, at $T=30$~K, an inelastic peak
centered around 12~meV is observed in constant-$Q$ scans at the 1-D AF
zone center $(1.5,0,0)$ [Fig.~\ref{exdata1}(a)]. Careful measurements at
different wave vectors have shown that this excitation is virtually
identical to Haldane gap excitations in Y$_{2}$BaNiO$_{5}$: (1) The
dispersion is very steep along the chain direction, and has minima at
$Q_{\|}=Q_{\|}^{(n)}\equiv (2n +1)\pi a^{\ast}$, where the subscript
``${\|}$'' indicates a projection on the $a$ axis and $n$ is integer.
Around its minima the dispersion is parabolic and can be approximated as
$(\hbar
\omega_{{\bf Q}})=\Delta^{2}+c_{0}^{2}(Q_{\|}-Q_{\|}^{(n)})^{2}$, where
$\Delta$ is the spin gap energy, and $c_{0}$ is the spin wave velocity.
A complete 3-axis deconvolution analysis of a set of constant-$Q$ scans
yielded $\Delta=10.4(0.1)$~meV and $c_{0}=200(11)$~meV \AA~ at $T=30$~K.
We see that the gap is very similar to that in the Y-nickelate.
Moreover, knowing $c_{0}$ allows us to determine the in-chain exchange
constant $J=c_{0}/(2.7 a)\approx 223$~K, also close to $J=285$~K in
Y$_{2}$BaNiO$_{5}$. (2) The transverse dispersion is almost zero. In
fact, it is at least an order of magnitude too small to induce any
long-range order through softening of Haldane excitations at the 3-D AF
zone center,\cite{Affleck89-2} as is the case in CsNiCl$_3$, for
example. (3) The energy-integrated intensity of the excitation decreases
rapidly as one moves away from the 1-D AF zone centers. The measured
dynamic structure factor $S({\bf Q},w)$ can be accurately fit
[Fig.~\ref{exdata1}, solid lines] to the double Lorntzian (DL) form
(Eq.~3 in Ref.~\cite{Regnault94}), predicted for Haldane-gap
excitations. The fitting procedure yields a value for the in-chain
correlation length $\xi$ which, to within experimental errors,
satisfies the relation $\Delta
\approx c_{0}/\xi$, expected for Haldane-gap
systems \cite{Nomura89,Golinelli92}. (4) Measurements  in different
Brillouin zones have shown that  the energy-integrated intensity of the
inelastic peak depends only on $Q_{\|}$. The dynamic structure factor is
purely 1-D and shows no 3-D intensity modulation expected from Ni-Pr
interference. Thus the gap excitations propagate {\em exclusively} on
the Ni-sites, without any involvement of the Pr spins. In other words,
as  long as the system is in the paramagnetic phase, the dynamics of the
Ni-spins do not depend on the properties of the $R$-substitute, and it
is natural to assume that in Pr$_{2}$BaNiO$_{5}$, just as in
Y$_{2}$BaNiO$_{5}$, we are looking at real Haldane-gap excitations
propagating on the Ni$^{2+}$ chains. The most exciting result is that
below $T_{N}$, i.e., in the magnetically ordered state, the Ni-chain
excitations survive [Fig.~\ref{exdata1}(b)]. As will be discussed in
detail below, the gap energy increases, while the intensity is
decreased. The point to be emphasized here is that the change in
excitation energy is {\em not} due to an increase  in the transverse
dispersion of the gap modes,  as in CsNiCl$_{3}$\cite{Buyers86,Morra88}:
the whole dispersion surace goes up, {\it independently} of the momentum
transfer perpendicular to the chain direction.  In fact, {\it all} the
1-D features discussed in the previous paragraph, including the
Ni-chain-only structure factor are preserved. Conventional acoustic spin
waves, i.e. Goldstone modes associated with long-range magnetic order,
were also observed: they are separate entities and coexist with the
Ni-chain gap modes. Unlike the latter however, the acoustic spin waves
represent correlated spin fluctuations of {\em both} Ni and $R$ moments.
Their intensity varies significantly from one 3-D Brillouin zone to the
next, and the overall intensity pattern follows that of the magnetic
Bragg reflections.

We now turn to discussing the results obtained for
(Nd$_{x}$Y$_{1-x}$)$_2$BaNiO$_5$ \cite{Zheludev96NBANO,Yokoo97NDY}.
While no single-crystal data is available for these materials at the
moment, inelastic neutron scattering experiments on powder samples have
provided detailed information on the temperature dependence of the
energy gap and on the effect of Y-substitution. For $x=1, 0.75, 0.5,$
and $0.25$ the compounds order magnetically with $T_{N}=48, 39, 29.5, $
and $19$~K, respectively. The saturation moment of the Ni-sublattice in
all cases with $x\ge 0.5$ is about $1.1 \mu_{B}$.  Observing the Haldane
gap modes in powder samples is quite possible,  thanks to the
1-dimensionality of the system. A typical constant-energy scan taken at
the gap energy $\Delta=10$~meV in (Nd$_{0.5}$Y$_{0.5}$)$_2$BaNiO$_5$ is
shown in Fig.~\ref{exdata2}(a) and has the characteristic saw-tooth
shape produced when the  DL 1-D structure factor is averaged over the
direction of the scattering vector ${\bf Q}$
\cite{Darriet93,Zheludev96NBANO} [solid lines in Fig.~\ref{exdata2}]. At
$|Q|>Q_{\|}^{0}$ a peak in constant-$Q$ scans is always observed at the
gap energy $\Delta$ [Fig.~\ref{exdata2}(b)].

By measuring the temperature dependence of the gap energy in samples
with different Y-content and N\'{e}el temperatures we can finally
observe, step by step, how the transition from the purely 1-D
quantum-disordered case of Y$_{2}$BaNiO$_{5}$ to the magnetically
ordered case of Nd$_{2}$BaNiO$_{5}$ occurs. Figure~\ref{vst}(a)
summarizes the temperature evolution of the gap energy in  several
(Nd$_{x}$Y$_{1-x}$)$_2$BaNiO$_5$ samples, including the pure Y-nickelate
($x=0$) \cite{Sakaguchi96}. One readily sees the ``universal'' features:
(1) In the paramagnetic phase, as far as the Ni-chain modes are
concerned, there is virtually no difference between samples with
different $R\ne Y$ content. The gradual increase of the gap energy with
temperature is a well-known intrinsic   property of the 1-D $S=1$
Heisenberg model \cite{Golinelli94}, and has been previously seen in
several Haldane-gap compounds experimentally
\cite{Ma95,Zheludev96-NINAZ}, including Y$_{2}$BaNiO$_{5}$
\cite{Sakaguchi96}. (2) In the ordered phase the gap starts to increase,
and the increase is linear with $(T_{N}-T)$. This is a totally new
effect that  has also been found in single-crystal measurements on
Pr$_{2}$BaNiO$_{5}$ \cite{Zheludev96PBANO}. (3) The temperature
dependence of the energy-integrated intensity in the gap excitations
also shows some universal features: it decreases at  high and low
temperatures, extrapolates to a finite value at $T=0$ and is always a
maximum at $T\approx T_{N}$ \cite{Yokoo97NDY}.

How can we account for the presence of Haldane gap excitations
propagating on the Ni-chains in a system that is not only magnetically
ordered in 3 dimensions, but where the ordered moment on the chain sites
is substantial? We suggest that in $R_2$BaNiO$_5$ the inter-chain
interactions are not sufficiently strong to destroy the Haldane singlet
ground state of the individual $S=1$ chains. Magnetic ordering can be
understood if one considers  singlet-ground-state  Ni-chains that get
{\em polarized} by an effective staggered exchange field produced by the
ordered $R$-sublattice \cite{Zheludev96PBANO,Zheludev96NBANO}. Indeed,
for a $S=1$ 1-D AF, the dynamic structure factor at $Q_{\|}=\pi
a^{\ast}$ is non-zero, the static staggered susceptibility is also
finite, and hence any staggered field will induce a finite staggered
magnetization. Of course, since both Ni and $R$ are essential for
establishing a 3-D spin network \cite{Zheludev96PBANO}, the problem is
to be treated self-consistently, which will automatically explain why
the Ni and $R$-spins order simultaneously.  Interestingly, this simple
scenario can also explain the increase of the Haldane gap energy in the
ordered phase. Recently S. Maslov \cite{Maslov97} has performed a
rigorous theoretical analysis of our  ``staggered field'' model and
shown that the change in the gap energy below $T_{N}$ should be
quadratic with the staggered magnetization induced on the Ni-chains.
Lets assume that the temperature dependence of the Ni moments is a
function of $T/T_{N}$ alone. This is a crude approximation, yet it is
consistent with some preliminary data \cite{Yokoo97NDY}. Noting that the
saturation Ni moment in Pr$_2$BaNiO$_5$ and
(Nd$_{x}$Y$_{1-x}$)$_2$BaNiO$_5$ (except $x=0.25$, where it has not yet
been measured) are almost equal, we can expect the change in $\Delta$
below $T_{N}$ to be a universal function of $(T/T_{N})$ in {\em all}
these systems. This  conjecture is indeed consistent with experiment:
Fig.~\ref{vst}(b) shows the increase of the spin gap in
(Nd$_{x}$Y$_{1-x}$)$_2$BaNiO$_5$ and Pr$_2$BaNiO$_5$ compared to that in
Y$_2$BaNiO$_5$ plotted  against $(T/T_{N})$. We see that all the
experimental points fall on a single master curve. More accurate
measurements of the $T$-dependencies of the sublattice magnetizations
and gap energies in (Nd$_{x}$Y$_{1-x}$)$_2$BaNiO$_5$, Pr$_2$BaNiO$_5$
and other systems will in the future enable us to draw more quantitative
conclusions.

In conclusion, a large amount of experimental data on magnetic
excitations in $R_2$BaNiO$_5$ systems has been accumulated so far. A
consistent theoretical understanding of the remarkable coexistence of
long-range order and Haldane-gap excitations is emerging. Nevertheless,
many unanswered questions still remain and further work is required. A
top priority is to verity the triplet nature of the Ni-chain excitations
and measure the temperature dependence of the gap separately for each
component. Conventional spin waves in $R_2$BaNiO$_5$ compounds should be
studied in greater detail. Systems like Er$_2$BaNiO$_5$ should also be
investigated, since they have different magnetic structures and may
exhibit different behavior. Finally, if the model presented above is
correct, in isomorphous {\em half-integer}-spin systems like
Nd$_2$BaCoO$_{5}$ \cite{Velasco96} the gap excitations should be totally
absent, but only preliminary results exist now to confirm this
conjecture \cite{Raymond96}.

It  is a pleasure to acknowledge the invaluable contributions  made by
my collaborators J. M. Tranquada, S. Maslov, J. Hill, T. Vogt and S.
Raymond (BNL), D. Buttrey  (University of Delaware), and T. Yokoo, M.
Nakamura and J. Akimitsu (Aoyama-Gakuin University, Tokyo, Japan). Work
at Brookhaven National Laboratory and experiments at the High Flux Beam
Reactor were carried out under Contract No. DE-AC02-76CH00016, Division
of Material Science, U.S.\ Department of Energy, as well as the
U.S.-Japan Cooperative Program on Neutron Scattering. The Aoyama-Gakuin
group was partially supported by a Grant-in-Aid for Scientific Research
from the Ministry of Education, Science and Culture Japan and The
Science Research Fund of Japan Private School Promotion Foundation.


\pagebreak

\begin{figure}
\caption{Example constant-$Q$ scans showing the Ni-chain gap excitations in
single-crystal Pr$_{2}$BaNiO$_{5}$ measured above (a) and below (b) the
N\'{e}el temperature $T_{N}=24$~K (from
Ref.~\protect\cite{Zheludev96PBANO}).}
\label{exdata1}
\end{figure}

\begin{figure}
\caption{Example constant-$E$ (a) and constant-$Q$ (b) scans showing the Ni-chain
gap excitations in NdYBaNiO$_{5}$. The measurements were done on a
powder sample above the magnetic ordering temperature $T_{N}=29.5$~K
(from Ref.~\protect\cite{Yokoo97NDY}).}
\label{exdata2}
\end{figure}

\begin{figure}
\caption{a) Temperature dependence of the Haldane gap energy measured in several
(Nd$_{x}$Y$_{1-x}$)$_2$BaNiO$_5$ powder samples (data from
Refs.~\protect\cite{Zheludev96NBANO,Yokoo97NDY}). b) Haldane gap
energies in several (Nd$_{x}$Y$_{1-x}$)$_2$BaNiO$_5$ and
Pr$_{2}$BaNiO$_{5}$ compounds plotted against $T/T_{N}$. The data are
taken from
Refs.~\protect\cite{Zheludev96PBANO,Zheludev96NBANO,Yokoo97NDY}.}
\label{vst}
\end{figure}

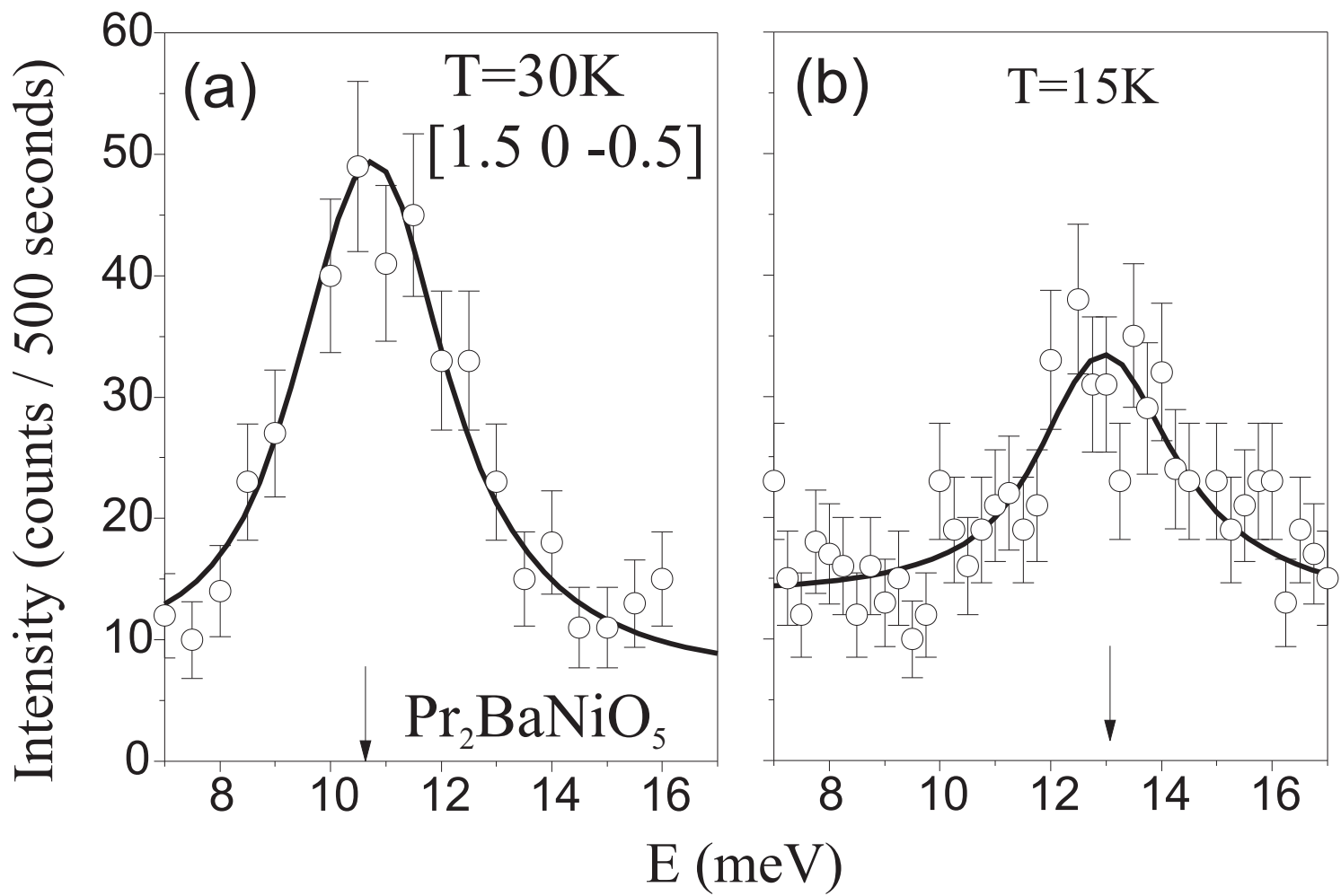
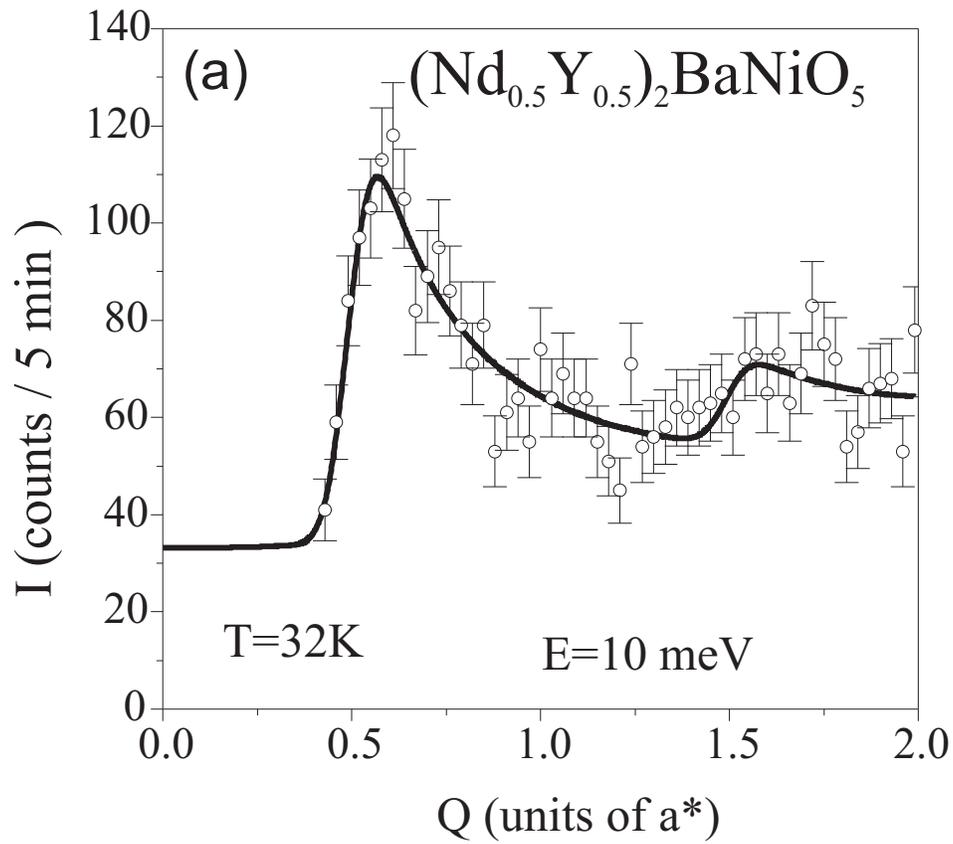
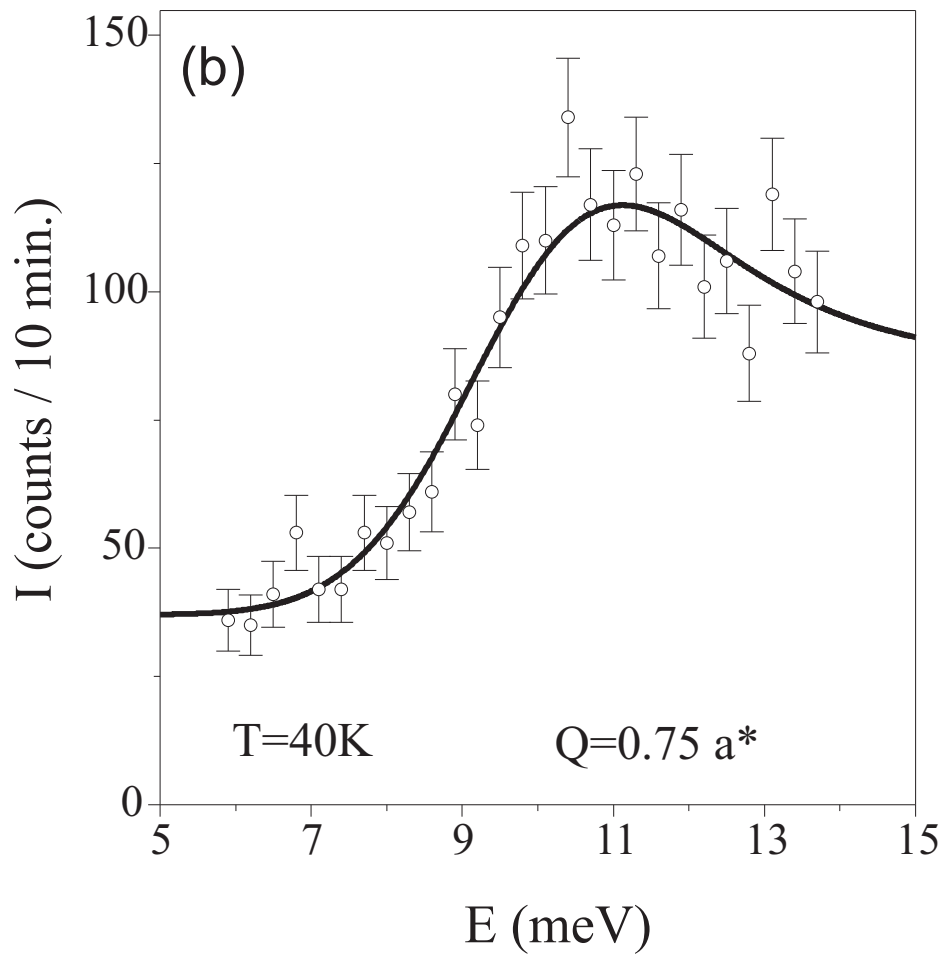
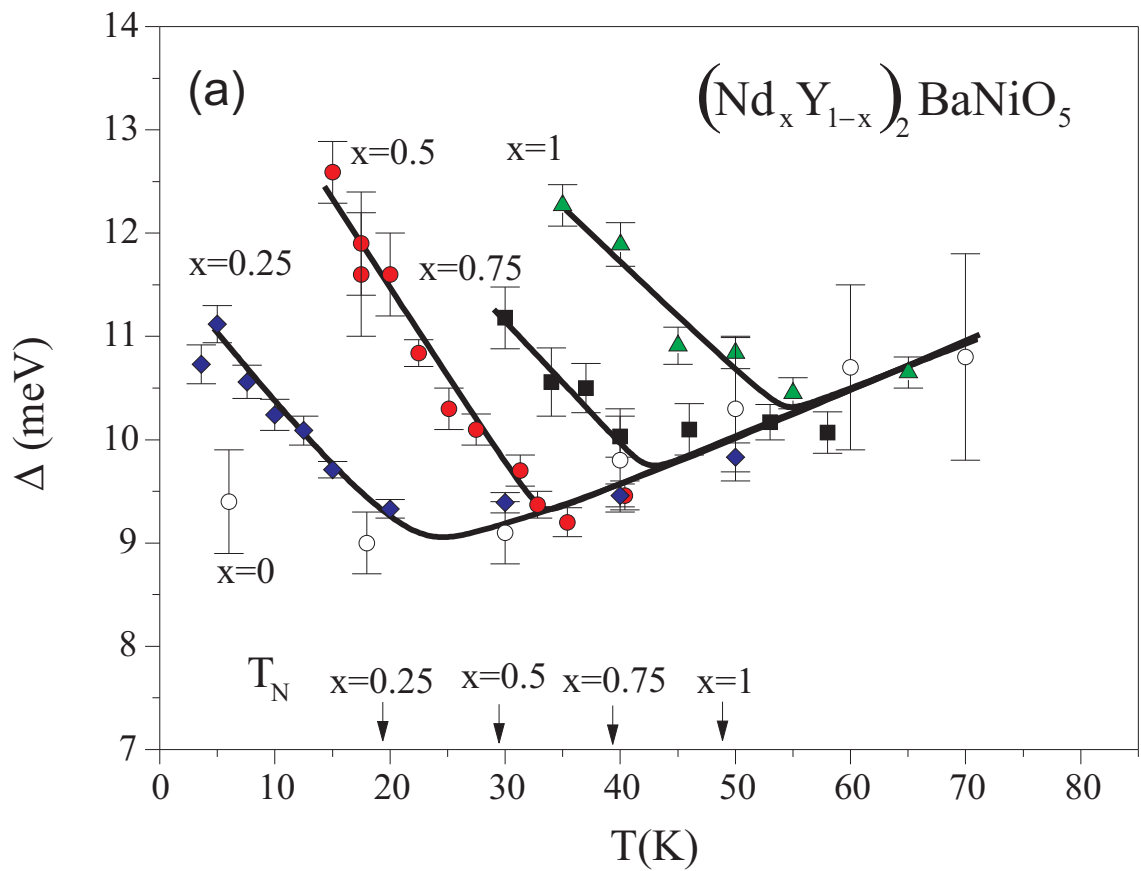
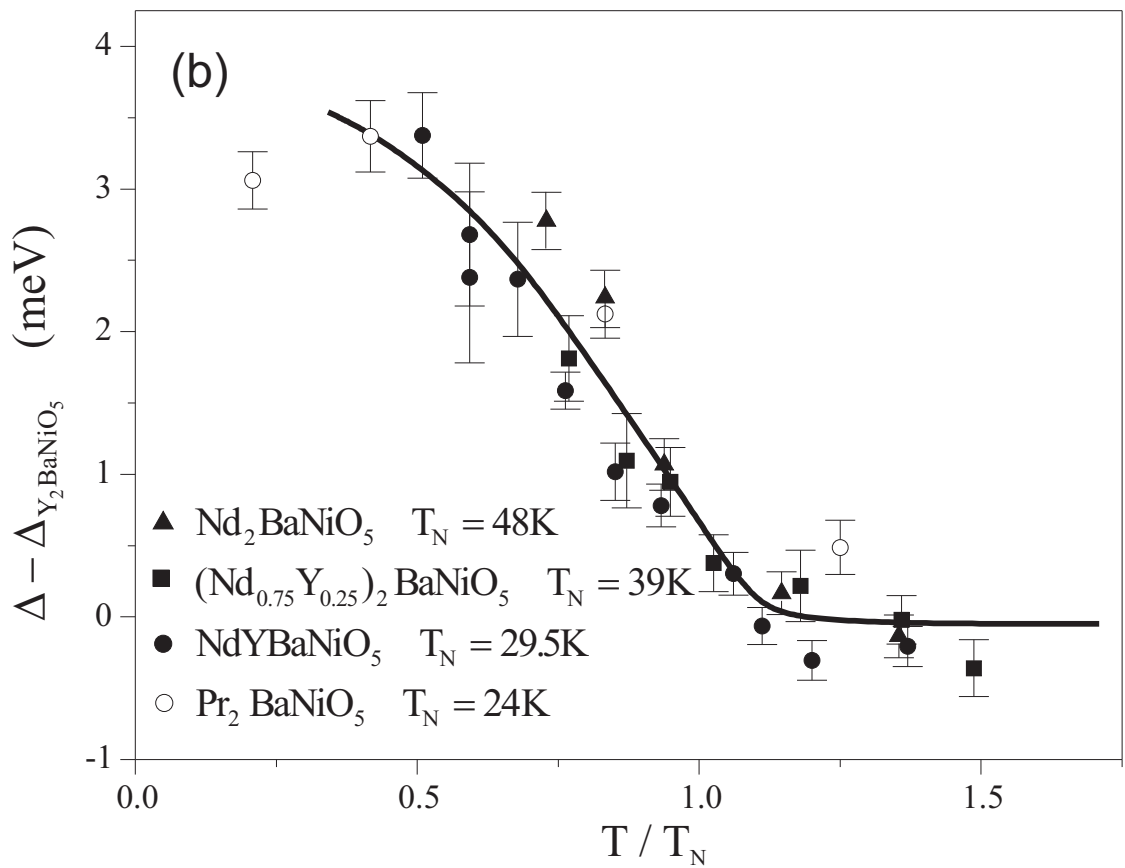
\end{document}